\documentclass[
showpacs,preprintnumbers,
nofootinbib,
 amsmath,amssymb,
 aps,
prd,
twocolumn
]{revtex4-1}

\usepackage[utf8]{inputenc}
\usepackage{epsf}
\usepackage{graphicx} 
\usepackage{amssymb}
\usepackage{dcolumn}
\usepackage{amsmath}
\usepackage{hyperref}
\usepackage{multirow}
\usepackage{booktabs}
\usepackage{siunitx}
\usepackage{soul}
\usepackage{color}
\newcommand{\kcmb}{\kappa_{\rm cmb}}
\newcommand{\kgal}{\kappa_{\rm gal}}

\def\ellB{\mbox{\boldmath$\ell$}}

\def\lsim{~\rlap{$<$}{\lower 1.0ex\hbox{$\sim$}}}
\def\gsim{~\rlap{$>$}{\lower 1.0ex\hbox{$\sim$}}}
\def\CB{\mbox{\boldmath${\rm C}$}}
\def\kg{\kappa_{\rm gal}}
\def\kcmb{\kappa_{\rm cmb}}

\def\cl{C_\ell}

\begin{document}

\title{Constraining Multiplicative Bias in CFHTLenS Weak Lensing Shear Data}
\author{Jia Liu}
 \email{jia@astro.columbia.edu}
 \author{Alvaro Ortiz-Vazquez}
\author{J. Colin Hill}
\affiliation{Department of Astronomy and Astrophysics,\\ Columbia University, New York, NY 10027, USA
}
\date{\today}
\begin{abstract}
Several recent cosmological analyses have found tension between constraints derived from the Canada-France-Hawaii Telescope Lensing Survey (CFHTLenS) data and those derived from other data sets, such as the \emph{Planck} cosmic microwave background (CMB) temperature anisotropies.  Similarly, a direct cross-correlation of the CFHTLenS data with \emph{Planck} CMB lensing data yielded an anomalously low amplitude compared to expectations based on \emph{Planck} or WMAP-derived cosmological parameters (Liu \& Hill 2015).  One potential explanation for these results is a multiplicative bias afflicting the CFHTLenS galaxy shape measurements, from which shears are inferred.  Simulations are used in the CFHTLenS pipeline to calibrate such biases, but no data-driven constraints have been presented to date.  In this paper, we cross-correlate CFHTLenS galaxy density maps with CFHTLenS shear maps and \emph{Planck} CMB lensing maps to calibrate an additional multiplicative shear bias ($m$) in CFHTLenS (beyond the multiplicative correction that has already been applied to the CFHTLenS galaxy shears), following methods suggested by Vallinotto (2012) and Das et al.\ (2013).  We analyze three magnitude-limited galaxy samples, finding $2$--$4\sigma$ evidence for $m<1$ using the deepest sample ($i < 24$), while the others are consistent with $m=1$ (no bias). This matches the expectation that the shapes of faint galaxies are the most prone to measurement biases.  Our results for $m$ are essentially independent of the assumed cosmology, and only weakly sensitive to assumptions about the galaxy bias. We consider three galaxy bias models, finding in all cases that the best-fit multiplicative shear bias is less than unity (neglecting photometric redshift errors and intrinsic alignment contamination).  A value of $m \approx 0.9$ would suffice to reconcile the amplitude of density fluctuations inferred from the CFHTLenS shear two-point statistics with that inferred from \emph{Planck} CMB temperature data.  This scenario is consistent with our results.
\end{abstract}
\maketitle

\section{Introduction}

Weak gravitational lensing occurs when the large-scale structure (LSS) of the universe distorts the path of light rays from a background source (a galaxy or the cosmic microwave background, CMB). It is a promising tool to probe the nature of dark energy, the total mass of neutrinos, and possible deviations from general relativity. Large galaxy lensing datasets, such as the ones from the Large Synoptic Survey Telescope~\cite{LSSTSciBook2009} and the Euclid Space Mission~\cite{Euclid2009}, will come online in the next decade. While providing unprecedentedly precise measurements of the LSS, these surveys also present a great challenge, as measurement systematics must be minimized in order to realize the surveys' full statistical power.

Major known galaxy lensing systematics include galaxy shape (or ``shear'') measurement  errors, photometric redshift calibrations, and intrinsic alignments of galaxies. In this work, we study the impact of one type of shape measurement systematic, the multiplicative bias, in the first large galaxy lensing survey --- the 154~deg$^2$  Canada-France-Hawaii Telescope Lensing Survey~(CFHTLenS)~\cite{Heymans2012}. The multiplicative bias originates from the mismatch of galaxy shapes assumed in image analysis models and those of real galaxies and/or from the non-linear relationship between image pixels and galaxy shape \cite{VoigtBridle2010, Kacprzak2014}, and is more likely to occur for faint galaxies. The multiplicative bias can change the overall amplitude of the cosmic shear auto-correlation and its cross-correlation with other probes of the LSS, hence causing a biased estimation of cosmological parameters. 
Ref.~\cite{Miller2013} details the procedure taken by the CFHTLenS team to calibrate the multiplicative bias, $m$, using the GREAT and SkyMaker simulations, where $m$ is fit as a function of signal-to-noise ratio and galaxy size. The resulting correction applied to the actual CFHTLenS shear measurements is $\approx$ 5--10\%, with larger (smaller) corrections for lower (higher) signal-to-noise galaxies. High-quality, all-sky CMB lensing data from \emph{Planck} have become public since the CFHTLenS data were published, allowing new data-driven constraints on the multiplicative bias, without the necessity of relying on galaxy image simulations~\cite{Vallinotto2012,Das2013}. 

Mild discrepancies between cosmological parameters estimated using galaxy lensing data and those estimated from CMB temperature measurements have been reported by several groups~\cite{Planck2015XIII, Hand2015, LiuHill2015, Raveri2015, Grandis2015}. For example, the cosmological parameter $\sigma_8(\Omega_m/0.27)^{0.46}$, which is orthogonal to the $\Omega_m$-$\sigma_8$ degeneracy direction for galaxy lensing, is lower by $\approx 2$--$2.5\sigma$ when estimated from CFHTLenS cosmic shear two-point statistics than when estimated from \emph{Planck} CMB temperature measurements~\cite{Planck2013XVI, Planck2015XIII, Heymans2013, Benjamin2013}. Here, $\sigma_8$ is the rms amplitude of linear density fluctuations on $8$ Mpc$/h$ scales at redshift zero.  Such a disagreement can potentially be explained by a multiplicative shear bias $m<1$, where $m=1$ corresponds to no bias. In this paper, we estimate $m$ through a joint analysis of the cross-correlations of (1) maps of galaxy number density and galaxy lensing convergence, and (2) maps of galaxy number density and CMB lensing convergence.

The paper is organized as follows. We first introduce the formalism in Sec.~\ref{formalism} and our data analysis procedures in Sec.~\ref{data}. We then present our results in Sec.~\ref{results} and discuss the implications in Sec.~\ref{discuss}.

\section{Formalism}\label{formalism}

In the Limber approximation~\cite{Limber1954}, the angular cross-power spectrum of two different probes (denoted $\alpha$ and $\beta$) of the LSS can be expressed in general as
\begin{eqnarray}
C_\ell^{\alpha\beta} = \int_0^\infty \frac{dz}{c} \frac{H(z)}{\chi^2(z)} W^\alpha(z)W^\beta(z) P\left(k = \frac{\ell}{\chi(z)}, z \right)
\end{eqnarray}
where $z$ is the redshift, $c$ is the speed of light, $H(z)$ is the Hubble parameter, $\chi(z)$ is the comoving distance, and $P(k, z)$ is the matter power spectrum at redshift $z$ and wavenumber $k$.
Assuming a flat universe, the weighting kernels $W(z)$ for galaxy lensing convergence ($\kg$), CMB lensing convergence ($\kcmb$), and galaxy number density~($\Sigma$) are
\begin{eqnarray}
W^{\kgal}(z) &=& \frac{3}{2}\Omega_{m} H_0^2 \frac{(1+z)}{H(z)} \frac{\chi(z)}{c}\\ \nonumber
&\times& \int_z^{\infty} dz_s \frac{dn(z_s)}{dz_s} \frac{\chi(z_s) - \chi(z)}{\chi(z_s)},\\
W^{\kcmb}(z) &=&  \frac{3}{2}\Omega_{m}H_0^2  \frac{(1+z)}{H(z)} \frac{\chi(z)}{c}  \frac{\chi(z_\star)-\chi(z)}{\chi(z_\star)},\\
W^{\Sigma}(z) &= & b(z)\frac{dn(z)}{dz} ,
\end{eqnarray}
where $\Omega_m$ is the matter density (relative to critical) at $z=0$, $H_0=H(z=0)$, $z_s$ is the redshift of the background source, where $z_\star=1100$ for the CMB, and $b(z)$ is the galaxy bias.  We neglect possible scale-dependence of the galaxy bias, as the moderate signal-to-noise ratio of our $\cl^{\kcmb\Sigma}$ measurement (see below) does not permit strong constraints on extended models.

The multiplicative bias can be estimated using a combination of auto- and cross-correlations involving $\kg$, $\kcmb$, and $\Sigma$ (see discussions in \cite{Vallinotto2012, Das2013}).  In this work, we use cross-correlations, 
\begin{eqnarray}
\label{eq.biasdef}
C_\ell^{\kg\Sigma,{\rm obs}} &=& m C_\ell^{\kg\Sigma,{\rm theory}}(b) \\
C_\ell^{\kcmb\Sigma,{\rm obs}} &=& C_\ell^{\kcmb\Sigma,{\rm theory}}(b), 
\end{eqnarray}  
to isolate the effect of $m$.  While $C_\ell^{\kg\Sigma,{\rm obs}}$ is sensitive to both $m$ and the galaxy bias $b$, $C_\ell^{\kcmb\Sigma,{\rm obs}}$ is sensitive to $b$ alone.  Thus, a joint analysis of these probes can break the degeneracy between $b$ and $m$, yielding robust constraints on both~\cite{Vallinotto2012, Das2013}.  The primary assumption of this method is that all data sets are governed by the same cosmological parameters (we assume minimal $\Lambda$CDM).  We also must make assumptions regarding the behavior of the galaxy bias $b(z)$, for which we consider three scenarios (see below).  Finally, we assume that the CMB lensing data are not afflicted by a multiplicative bias.

We use cosmological parameters obtained from \emph{Planck} 2015 data (TT, TE, EE + lowP, see Table 4 in Ref.~\cite{Planck2015XIII}). In particular, $\Omega_m = 0.3156$, $\sigma_8=0.831$, and $h=0.6727$.  We verify below that our results for $m$ are insensitive to the particular values assumed for these parameters.

\section{Data Analysis}\label{data}

We use the publicly available CMB lensing convergence ($\kcmb$) map released by the \emph{Planck} collaboration (2015 data release).  We use CFHTLenS data to construct $\kg$ and $\Sigma$ maps. The CFHTLenS survey consists of four sky patches located far from the Galactic plane (W1, W2, W3, and W4), with a total area of 154 deg${}^2$ and a limiting magnitude $i_{AB}<24.5$. The construction of the $\kcmb$ and $\kg$ maps is summarized in detail in Ref.~\cite{LiuHill2015}, with the only difference that we apply a redshift cut of $0.2<z<1.3$ to the CFHTLenS galaxy sample used in the $\kg$ reconstruction in this paper.  The effective number density of galaxies used in the $\kg$ reconstruction is 9.3~galaxies/arcmin$^{2}$.

It is important to note that we have already applied to the $\kg$ maps the multiplicative bias correction provided in the CFHTLenS catalogue \cite{Miller2013},
\begin{eqnarray}
\label{eq.mCFHT}
m_{\rm CFHT}(\nu_{\rm SN},r)=\frac{B}{\log_{10}(\nu_{\rm SN})}\exp(-A r \nu_{\rm SN}),
\end{eqnarray}
with $A=0.057$ and $B=-0.37$; $\nu_{\rm SN}$ is the signal-to-noise ratio and $r$ is the galaxy size. By their definition, the multiplicative bias vanishes when $1+m_{\rm CFHT}=1$, i.e., $m_{\rm CFHT} = 0$. Typical values of this correction are $1+m_{\rm CFHT} \approx 0.9$--$1$.  Any multiplicative bias detected in our work is in addition to this correction.  Recall that we define $m$ here such that $m=1$ corresponds to no bias --- e.g., see Eq.~(\ref{eq.biasdef}).  Also, in our work $m$ is an overall factor applied to the $\kgal$ map, whereas $m_{\rm CFHT}(\nu_{\rm SN},r)$ in Eq.~(\ref{eq.mCFHT}) is applied as an average of galaxies within the 1 arcmin smoothing scale (see Eq. 4 in ~\cite{Liu2015}).

We follow Ref.~\cite{Omori2015} to create $\Sigma$ maps, where three different magnitude cuts are applied to the galaxies: $18<i<22$, $18<i<23$, and $18<i<24$ (note that in comparison, we apply no magnitude cuts to the $\kg$ sample, other than the survey magnitude limit $i<24.5$), resulting in a mean redshift $\langle z \rangle=0.52$, 0.61, and 0.69, respectively.  For the $\Sigma$ maps, we include galaxies that have \texttt{lensfit} weight=0 (which are excluded from the $\kg$ sample) --- these objects are identified as galaxies, but they are too small to have shapes measured accurately for shear reconstruction.  Ref.~\cite{Choi2015} tested photo-$z$ errors with and without these galaxies and found no significant difference. When we exclude these galaxies (which account for 65\%, 55\%, and 45\% of the total galaxies for $i$$<$22, 23, 24 samples, respectively) in our analysis, the error bars increase by roughly a factor of 2, and hence we can draw no statistically significant conclusions regarding the multiplicative bias. 

The galaxy number density fluctuation $\Sigma_j$ in the $j^{\rm th}$ pixel on a grid map is calculated using
\begin{eqnarray}
N_j = \frac{N_{j,{\rm raw}}}{w_j},\\ 
\Sigma_j = \frac{N_j}{\left<N\right>}-1,
\end{eqnarray}
where $N_{j,{\rm raw}}$ is the number of galaxies falling within that pixel, and $w_j\in(0,1]$ is the unmasked fraction of that pixel calculated from degrading a high-resolution mask map. The galaxy number density is 3.3, 7.5, and 15.0~galaxies/arcmin$^{2}$ for the three galaxy samples (from shallowest to deepest).

\begin{figure}
\begin{center}
\includegraphics[width=0.5\textwidth]{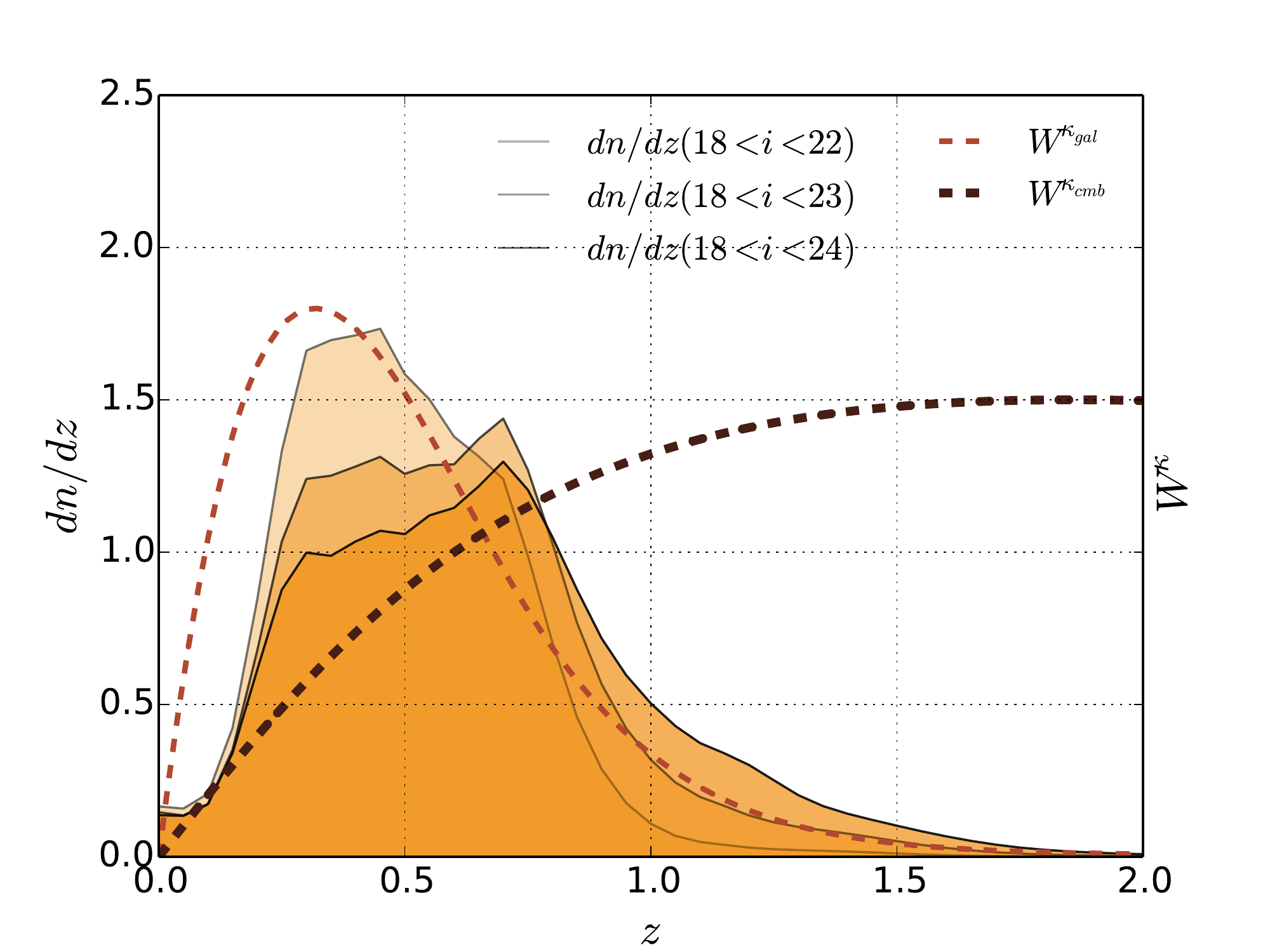}
\end{center}
\caption{\label{fig:kernel} The lensing kernels for CMB lensing (thick dashed) and CFHTLenS galaxy lensing (thin dashed) and the redshift distributions for the three galaxy number density samples considered in this work (solid shaded). All galaxies have a best-fit redshift $0.2<z<1.3$, and the full probability distribution of individual galaxies is used to calculate $dn/dz$ and $\kg$. The lensing kernels are rescaled here for display purposes only (the CMB lensing kernel is normalized to a maximum of 1.5 and the CFHTLenS kernel to a maximum of 1.8).}
\end{figure}

The galaxy redshift distributions and lensing kernels for $\kg$ (mean redshift $\langle z \rangle=0.74$) and $\kcmb$ are shown in Fig.~\ref{fig:kernel}. We use the publicly available masks provided by \emph{Planck} and CFHTLenS\footnote{We mask out pixels with \texttt{mask} $>$ 0 --- see Table B2 in Ref.~\cite{Erben2013} for a detailed description of the \texttt{mask} values.} and calculate the remaining sky fraction $f_{\rm sky}$ using the combination of these two masks, finding $f_{\rm sky}=0.00298$.

We estimate the two-dimensional (2D) auto- or cross-correlation via
\begin{eqnarray}
\label{eq: ps2d}
C^{\alpha \beta}(\ellB) = \hat M_\alpha(\ellB)^*\hat M_\beta(\ellB) \,,
\end{eqnarray}
where $\hat M_\alpha$ is the Fourier transform of the 2D map $M_\alpha$ ($\alpha, \beta \in [\kg,\kcmb, \Sigma]$), and $*$ denotes complex conjugation. We then average over pixels in each multipole bin, $|\ellB| \in (\ell-\Delta\ell/2, \, \ell+\Delta\ell/2)$, for five linearly spaced bins between $40\leq\ell\leq2000$.

We estimate parameters by minimizing
\begin{eqnarray}
\label{eq: chisq}
\chi^2 =\sum_{i,j} \left(O^i-N^i \right) \CB_{ij}^{-1} \left(O^j-N^j \right),
\end{eqnarray}
where the data vector $O = \left(\cl^{\kg \Sigma}, \cl^{\kcmb \Sigma} \right)$ contains 40 entries (2 cross-correlations, 4 CFHTLenS fields, each with 5 bins), and the model vector $N=N(b, m)$ is fixed at our base cosmology (\emph{Planck} 2015), with the galaxy bias and multiplicative shear bias as free parameters. The covariance matrix $\CB_{ij}$ is estimated using 100 realizations of $\kg$ maps, where we randomly rotate the galaxies\footnote{We note that the randomly rotated $\kg$ maps do not contain cosmic variance, and hence underestimate the variance in $\cl^{\kg\kg}$. 
However, the variance is dominated by galaxy shot noise for CFHTLenS. Moreover, the overall covariance $\CB_{ij}$ is dominated by the noise in the \emph{Planck} CMB lensing reconstruction. Therefore, the effect of omitting cosmic variance in the simulated $\kg$ maps is negligible.}, and 100 simulated \emph{Planck} CMB lensing maps. We apply a correction factor of $(n-p-2)/(n-1)$ to the inverse of the covariance matrix to obtain an unbiased estimator~\cite{Hartlap2007}, where $n=100$ and $p=40$ are the number of simulations and the number of bins. The diagonal components of $\CB_{ij}$ are consistent with the theoretical  Gaussian variance estimated from the auto-power spectra of the maps to within 10\%.

Because $b$ and $m$ are somewhat degenerate, we test the robustness of our $m$ constraints using three models for the galaxy bias: a constant $b$ and two redshift-dependent models, with $b(z)=b_0(1+z)$ (e.g.,~\cite{Ferraro2014}) or $b(z)=\tilde{b}_0(1+z)-z$~\cite{Tegmark1998}.  The last model is appropriate for tracers whose comoving number density is conserved after their formation at some early epoch.\footnote{This statement is only exact in an Einstein-de Sitter universe, but this does not restrict our phenomenological use of the model.}  Our constraints on $m$ and $b$ are given in the next section.

\section{Results}\label{results}

\begin{figure*}
\begin{center}
\includegraphics[width=\textwidth]{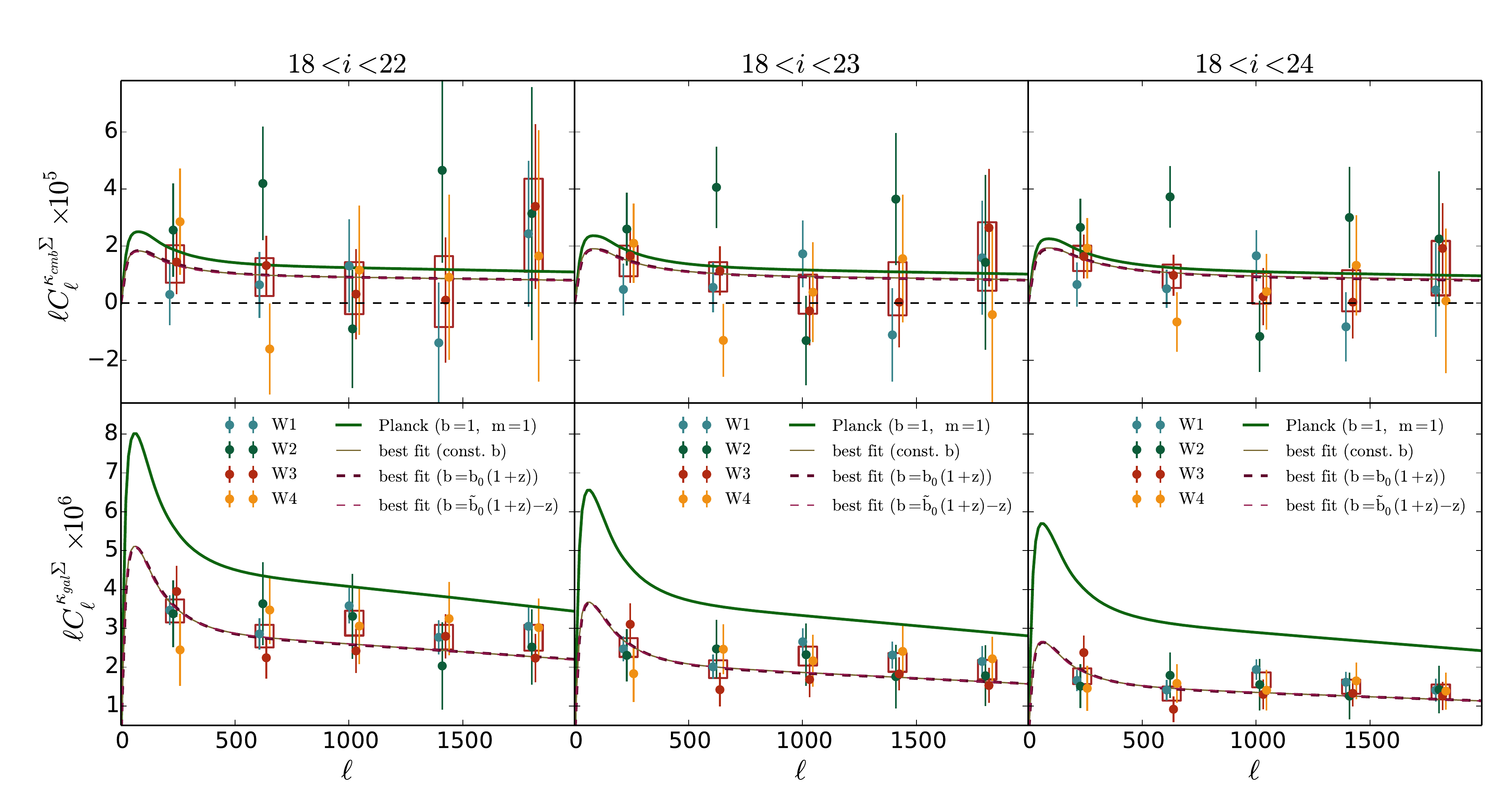}
\end{center}
\caption{\label{fig:CC} Angular cross-power spectra of (1) $\kcmb$ and $\Sigma$ (upper panels) and (2) $\kg$ and $\Sigma$ (lower panels) for three galaxy samples ($18<i<22$, $18<i<23$, and $18<i<24$). Data points are for individual CFHTLenS fields, and errors are estimated using 100 simulated $\kcmb$ maps and 100 randomly-rotated $\kg$ maps. The boxes represent the inverse-variance weighted sum of the four fields. The thick-solid, thin-solid, thick-dashed, and thin-dashed curves are the fiducial theoretical model using \emph{Planck} 2015 parameters ($b,m=1$), the best-fit model assuming a constant $b$, the best-fit model assuming $b(z)=b_0(1+z)$, and the best-fit model assuming $b(z)=\tilde{b}_0(1+z)-z$, respectively.  The three best-fit models use the combined constraints on $b$ and $m$ from jointly fitting the $\cl^{\kcmb\Sigma}$ and $\cl^{\kg\Sigma}$ data.} 
\end{figure*}

The cross-power spectra $\cl^{\kg\Sigma}$ and $\cl^{\kcmb\Sigma}$ for the three galaxy number density samples are shown in Fig.~\ref{fig:CC}, where we also overlay the fiducial theoretical models ($b=1, m=1$) and best-fit results.  For the best-fit models, we obtain $\chi^2 = 19.8$, 27.4,  24.1 for 38 degrees of freedom (corresponding to $p$-values of 0.994, 0.900, 0.961) for the $i$ $<$ 22, 23, 24 samples, respectively. The $\chi^2$ values are nearly identical for all three bias models, as the best-fit curves in each case are nearly indistinguishable (see Fig.~\ref{fig:CC}). The somewhat high $p$-values suggest that our error bars could be slightly overestimated, which could be due to the limited number of simulations used to determine the covariance matrix.

\begin{figure}
\begin{center}
\includegraphics[width=0.5\textwidth]{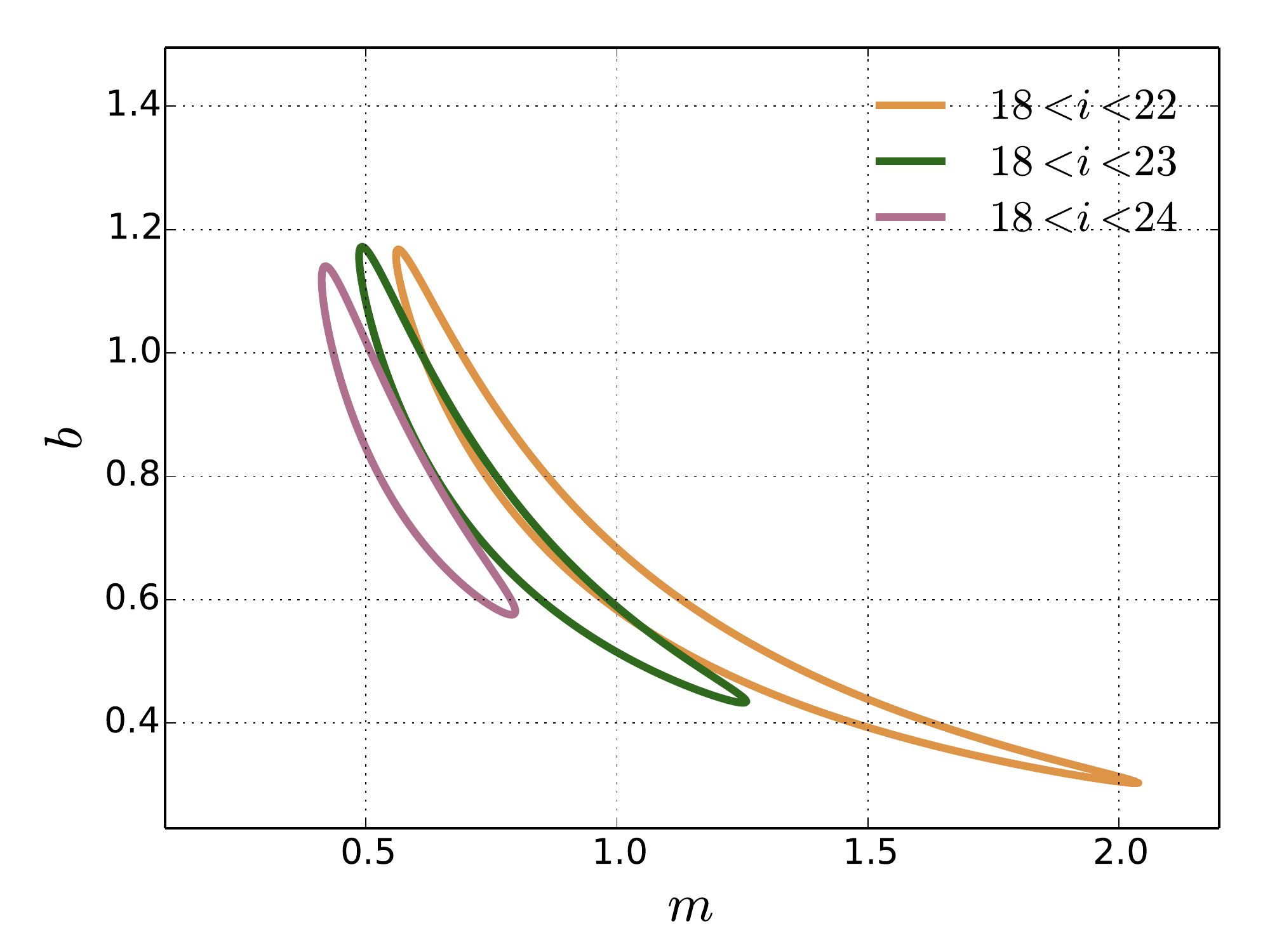}
\end{center}
\caption{\label{fig:contour_b} Error contours ($68$\%) in the $m$--$b$ plane, assuming a constant $b$.  The different contours correspond to different galaxy samples, as labeled. Marginalized values of $m$ and $b$ are listed in Table~\ref{tab:results_const_b}.  The deepest sample considered ($18<i<24$) shows evidence for a multiplicative bias $m<1$.}
\end{figure}

\begin{figure}
\begin{center}
\includegraphics[width=0.5\textwidth]{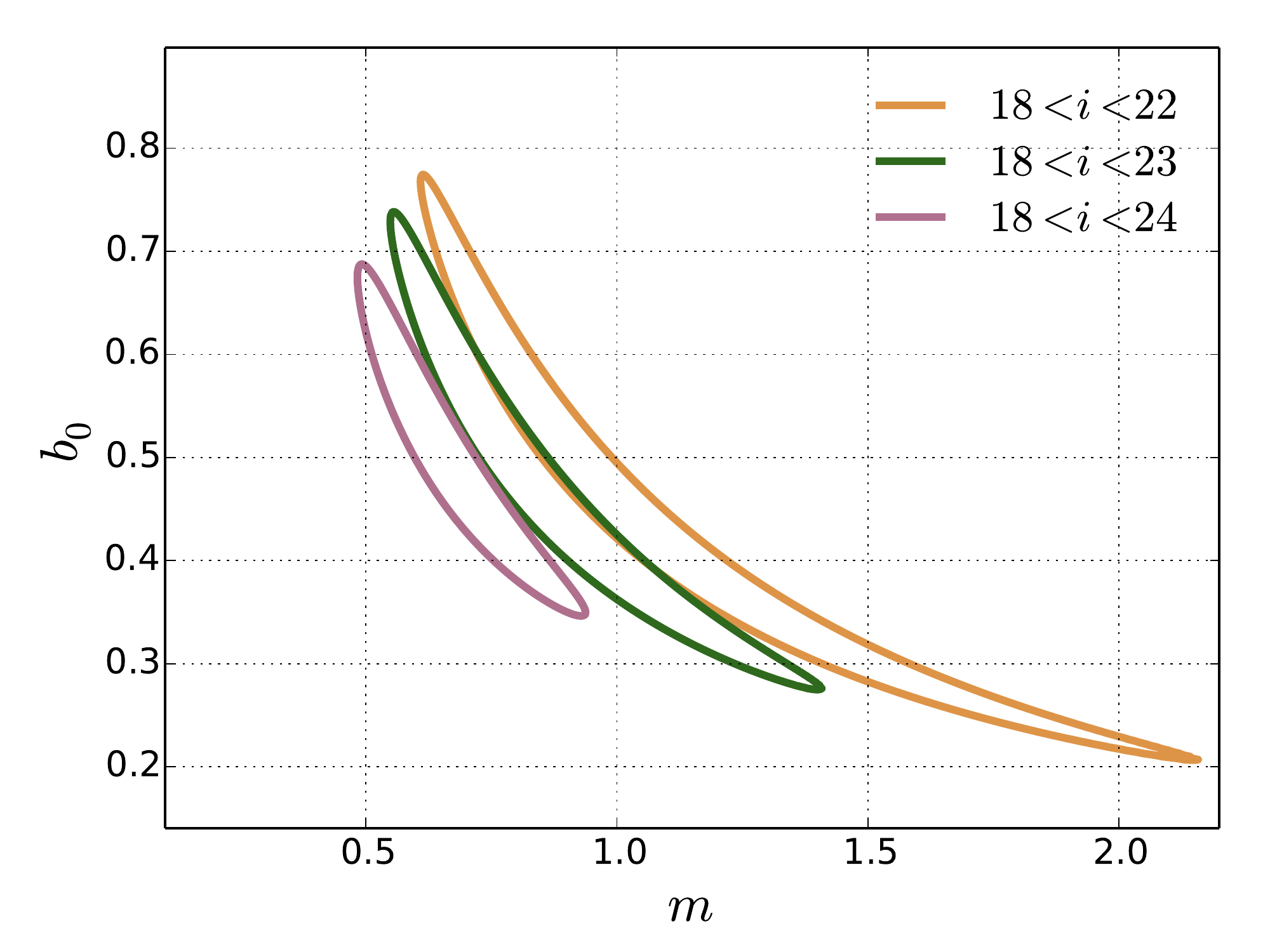}
\end{center}
\caption{\label{fig:contour_b0} Error contours (68\%) in the $m$--$b_0$ plane, assuming $b(z)=b_0(1+z)$. The different contours correspond to different galaxy samples, as labeled.  Marginalized values of $m$ and $b_0$ are listed in Table~\ref{tab:results_b0}.  As in Fig.~\ref{fig:contour_b}, the deepest sample considered ($18<i<24$) shows evidence for a multiplicative bias $m<1$.}
\end{figure}

\begin{figure}
\begin{center}
\includegraphics[width=0.5\textwidth]{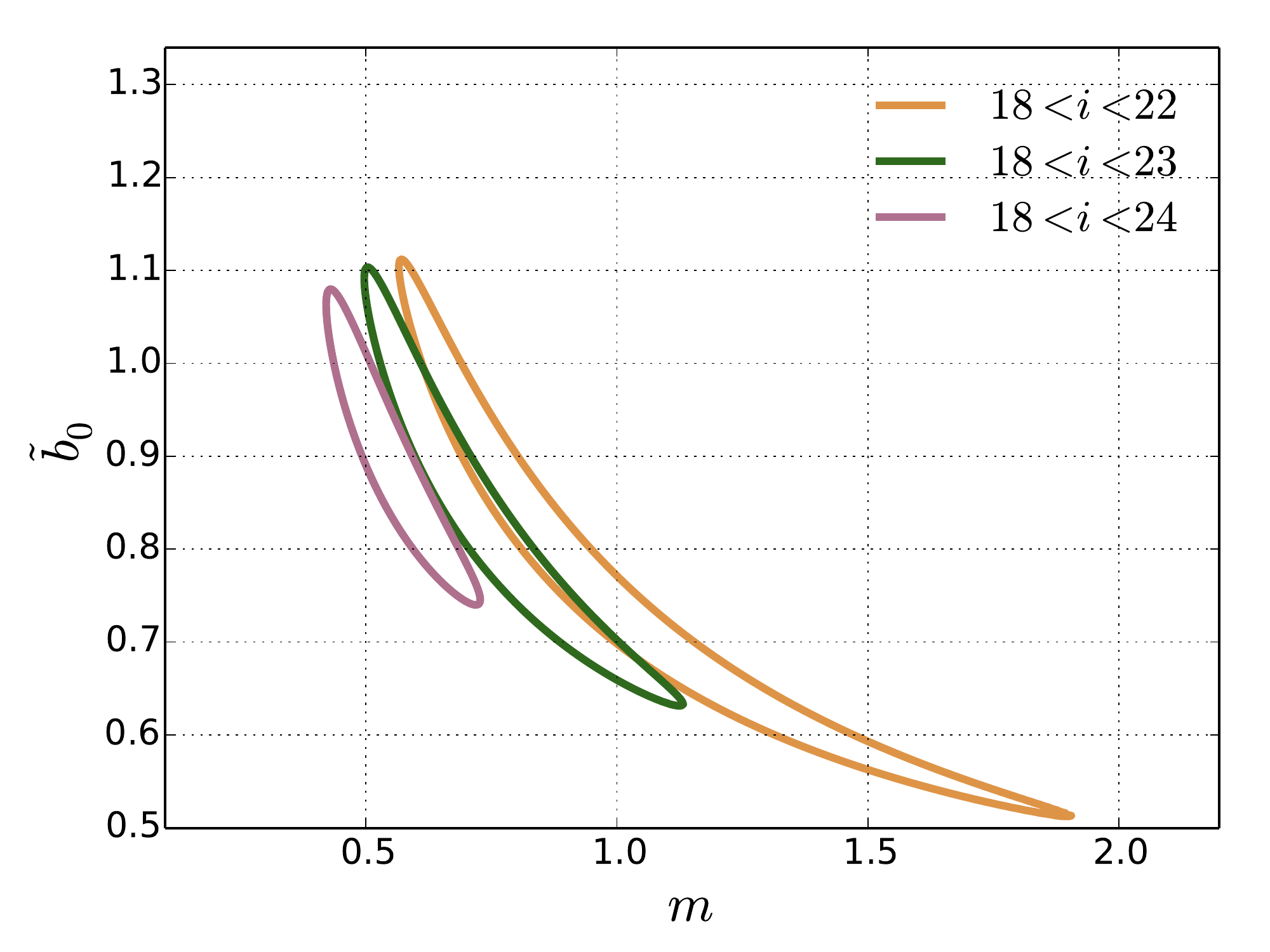}
\end{center}
\caption{\label{fig:contour_b0_2} Error contours (68\%) in the $m$--$\tilde{b}_0$ plane, assuming $b(z)=\tilde{b}_0(1+z)-z$. The different contours correspond to different galaxy samples, as labeled.  Marginalized values of $m$ and $b_0$ are listed in Table~\ref{tab:results_b0_2}.  As in Figs.~\ref{fig:contour_b} and~\ref{fig:contour_b0}, the deepest sample considered ($18<i<24$) shows evidence for a multiplicative bias $m<1$.}
\end{figure}

Fig.~\ref{fig:contour_b} shows the derived constraints on $b$ and $m$ from these two cross-correlations, assuming a constant $b$. Figs.~\ref{fig:contour_b0} and~\ref{fig:contour_b0_2} show the constraints for a redshift-dependent $b(z)=b_0(1+z)$ and $b(z)=\tilde{b}_0(1+z)-z$, respectively. The marginalized constraints are listed in Table~\ref{tab:results_const_b} (for a constant $b$), Table~\ref{tab:results_b0} (for $b(z)=b_0(1+z)$), and Table~\ref{tab:results_b0_2} (for $b(z)=\tilde{b}_0(1+z)-z$). In all of the tables, we list constraints on $b$ using $\cl^{\kcmb\Sigma}$ only and using $\cl^{\kg\Sigma}$ only (while assuming $m=1$), as well as joint constraints on $b$ and $m$ using the combination of these two cross-correlations.

\begin{table}[!h]
\begin{tabular}{c|c|c|c|cccccc}
\hline
\multirow{2}{*}{$\Sigma$ sample}		& 
$C_\ell^{\kcmb\Sigma}$ 		&  $C_\ell^{\kg\Sigma}$	& \multicolumn{2}{c}{$C_\ell^{\kcmb\Sigma}+C_\ell^{\kg\Sigma}$}	\\
\cline{2-5} & $b$	& $b$ ($m$=1)						& $b$					& $m$		\\
\hline	
$18<i<22$	&$0.78\substack{+0.26 \\ -0.26}	$		&$0.68\substack{+0.04 \\ -0.03}	$	&$0.60\substack{+0.26 \\ -0.28}$ &$0.77\substack{+0.53 \\ -0.22}$	\\
$18<i<23$ 	&$0.87\substack{+0.22 \\ -0.21}$		&$0.59\substack{+0.03 \\ -0.03}	$	&$0.73\substack{+0.24 \\ -0.24}$ &$0.65\substack{+0.30 \\ -0.15}$	\\
$18<i<24$ 	&$0.88\substack{+0.16 \\ -0.16}	$		&$0.49\substack{+0.03 \\ -0.02}	$	&$0.82\substack{+0.18 \\ -0.19}$ &$0.52\substack{+0.14 \\ -0.09}$	\\
\hline
\end{tabular}
\caption[]{\label{tab:results_const_b} Marginalized constraints on $b$ and $m$, where a constant $b$ is assumed. We provide constraints obtained using $\cl^{\kcmb\Sigma}$ only (column 2), $\cl^{\kg\Sigma}$ only (column 3), and their combination (columns 4 and 5).}
\end{table}

\begin{table}[!h]
\begin{tabular}{c|c|c|c|cccccc}
\hline
\multirow{2}{*}{$\Sigma$ sample}		& 
$C_\ell^{\kcmb\Sigma}$ 		&  $C_\ell^{\kg\Sigma}$ 	& \multicolumn{2}{c}{$C_\ell^{\kcmb\Sigma}+C_\ell^{\kg\Sigma}$}	\\
\cline{2-5}							& $b_0$	& $b_0$ ($m$=1)						& $b_0$					& $m$		\\
\hline															
$18<i<22$	&$0.52\substack{+0.17 \\ -0.17}	$		&$0.49\substack{+0.02 \\ -0.02}	$	&$0.40\substack{+0.17 \\ -0.18}$ &$0.83\substack{+0.55 \\ -0.23}$	\\
$18<i<23$ 	&$0.55\substack{+0.13 \\ -0.13}$		&$0.42\substack{+0.02 \\ -0.02}	$	&$0.46\substack{+0.15 \\ -0.15}$ &$0.73\substack{+0.33 \\ -0.17}$	\\
$18<i<24$ 	&$0.53\substack{+0.10 \\ -0.10}	$		&$0.35\substack{+0.02 \\ -0.02}	$	&$0.49\substack{+0.11 \\ -0.11}$  &$0.61\substack{+0.17 \\ -0.11}$	\\
\hline
\end{tabular}
\caption[]{\label{tab:results_b0} Marginalized constraints on $b_0$ and $m$, where $b(z)=b_0(1+z)$ is assumed for the behavior of the galaxy bias.  We provide constraints obtained using $\cl^{\kcmb\Sigma}$ only (column 2), $\cl^{\kg\Sigma}$ only (column 3), and their combination (columns 4 and 5).}
\end{table}

\begin{table}[!h]
\begin{tabular}{c|c|c|c|cccccc}
\hline
\multirow{2}{*}{$\Sigma$ sample}		& 
$C_\ell^{\kcmb\Sigma}$ 		&  $C_\ell^{\kg\Sigma}$ 	& \multicolumn{2}{c}{$C_\ell^{\kcmb\Sigma}+C_\ell^{\kg\Sigma}$}	\\
\cline{2-5}							& $\tilde{b}_0$	& $\tilde{b}_0$ ($m$=1)						& $\tilde{b}_0$					& $m$		\\
\hline															
$18<i<22$	&$0.85\substack{+0.17 \\ -0.17}	$		&$0.77\substack{+0.02 \\ -0.02}	$	&$0.73\substack{+0.18 \\ -0.20}$ &$0.77\substack{+0.49 \\ -0.21}$	\\
$18<i<23$ 	&$0.92\substack{+0.13 \\ -0.13}$		&$0.71\substack{+0.02 \\ -0.02}	$	&$0.83\substack{+0.15 \\ -0.16}$ &$0.65\substack{+0.24 \\ -0.14}$	\\
$18<i<24$ 	&$0.93\substack{+0.10 \\ -0.10}	$		&$0.64\substack{+0.02 \\ -0.02}	$	&$0.89\substack{+0.11 \\ -0.11}$ &$0.52\substack{+0.11 \\ -0.08}$	\\
\hline
\end{tabular}
\caption[]{\label{tab:results_b0_2}  Marginalized constraints on $b_0$ and $m$, where $b(z)=\tilde{b}_0(1+z)-z$ is assumed for the behavior of the galaxy bias.  We provide constraints obtained using $\cl^{\kcmb\Sigma}$ only (column 2), $\cl^{\kg\Sigma}$ only (column 3), and their combination (columns 4 and 5).}
\end{table}

From the $\cl^{\kcmb\Sigma}$-only and $\cl^{\kg\Sigma}$-only constraints in Table~\ref{tab:results_const_b}, it is apparent that the inferred galaxy bias is only clearly consistent for these two methods for the $i<22$ sample, with a marginal discrepancy seen for the $i<23$ sample and a non-negligible discrepancy seen for the $i<24$ sample. Moreover, while the $\cl^{\kcmb\Sigma}$-only measurements show an increasing galaxy bias as a function of $z$ (i.e., with increasing depth of the galaxy sample), the $\cl^{\kg\Sigma}$-only measurements show the opposite trend.  These results suggest that either a more complicated galaxy bias model is required or that one of the data sets is afflicted by a systematic.  Tables~\ref{tab:results_b0} and \ref{tab:results_b0_2} show the same trends, however, even when allowing for a redshift-dependent galaxy bias.  An obvious candidate explanation is thus a multiplicative shear bias afflicting $\kg$, which can be constrained in the joint analysis of the two cross-spectra.  

The joint analysis shows that $m$ is statistically consistent with unity (no bias) for the $i<22$ and $i<23$ samples, while we obtain $2$--$4\sigma$ evidence for $m<1$ using the $i < 24$ sample, depending on the galaxy bias model adopted. The $m$ constraints are statistically consistent for the three different galaxy bias models considered here. It is not surprising that $m<1$ is only significant for the deepest sample, as this cross-correlation probes the LSS at a higher redshift than the other two samples (see Fig.~\ref{fig:kernel}). At high redshifts, the $\kg$ signal receives more contributions from faint galaxies, whose shapes are more difficult to measure accurately.

We test the robustness of our constraints on $m$ to the assumed cosmological parameters by redoing the constant-$b$ analysis while using WMAP9 cosmological parameters (WMAP+eCMB+BAO+$H_0$ in Table 2 of Ref.~\cite{Hinshaw2013}), e.g., $h=0.697$, $\Omega_m=0.282$, and $\sigma_8=0.817$. Our multiplicative bias results are almost identical to those presented above (the change in the best-fit $m$ is $\lesssim 1$\% for all three galaxy samples), although the inferred galaxy bias values increase by $\approx 10\%$.  The evidence for $m<1$ is thus insensitive to the assumed cosmology.

Our measured auto-correlations of $\Sigma$ and cross-correlations of $\Sigma$ and $\kcmb$ are consistent with those presented in Ref.~\cite{Omori2015}, although the multipole bins used in the two analyses differ slightly. 

\section{Discussion}\label{discuss}

In this paper, we search for evidence of additional multiplicative biases in CFHTLenS weak gravitational lensing shear measurements (beyond the standard multiplicative correction from the CFHTLenS shear catalogue) using joint cross-correlations of CFHTLenS data and \emph{Planck} CMB lensing data. Our results show hints ($2$--$4\sigma$) of a non-vanishing multiplicative bias for the deepest sample of galaxies considered in this analysis. We stress that, despite our focus on biases in shear measurement, other systematics that can change the overall amplitude of $\cl^{\kg\Sigma}$ may also partially or even fully account for the discrepancy we see. Possible sources include intrinsic alignment contamination~\cite{Troxel2014,Chisari2015,Larsen2015} and photometric redshift errors~\cite{Hildebrandt2012,Benjamin2013,Choi2015,Kitching2016}, which are beyond the scope of this work, but must be studied more carefully in the future.  Another alternative would be an unexpectedly complex galaxy bias model --- a non-monotonic redshift dependence would be needed to explain the results in Tables~\ref{tab:results_const_b}--\ref{tab:results_b0_2}.

Our constraint on $m$ is somewhat degenerate with constraints on the galaxy bias $b$. To circumvent the additional uncertainty introduced by the modeling of the galaxy bias, one can limit the galaxy sample for $\Sigma$ to a thin redshift slice (preferably with spectroscopic redshift measurements), and hence $b(z)$ would be nearly the same for both cross-correlations ($\cl^{\kg\Sigma}$ and $\cl^{\kcmb\Sigma}$).  In this limit, any scale-dependence of the bias will also have a nearly identical effect on the two cross-correlations.  As a result, $m$ will be simply the ratio of $\cl^{\kg\Sigma}$ and $\cl^{\kcmb\Sigma}$ times a geometric factor (see Eq. 6 in Ref.~\cite{Das2013}). We have tested this idea using galaxies in the Sloan Digital Sky Survey (SDSS). However, the low number density of galaxies in the SDSS sample ($<$0.05~galaxy/arcmin$^2$, compared with $\approx$10~galaxy/arcmin$^2$ in CFHTLenS) is insufficient to obtain statistically significant constraints from the cross-correlations within the CFHTLenS sky area.

To place this work in context, we estimate the level of multiplicative bias needed to reconcile the tension between cosmological parameter constraints derived from CFHTLenS two-point statistics and those derived from \emph{Planck} CMB temperature anisotropy measurements.  We use the fact that the auto-power spectrum of $\kg$ scales roughly quadratically with $\sigma_8$ and exactly quadratically with $m$. Ref.~\cite{Planck2013XVI} found that $\sigma_8(\Omega_m/0.27)^{0.46}=0.89\pm0.03$ (using ``\emph{Planck}+WP+highL'' data), compared with $0.774\pm0.04$ from CFHTLenS~\cite{Heymans2013}. Therefore, a multiplicative bias $m\approx0.9$ suffices to bridge the gap between these two measurements. Such a bias would also help reconcile the discrepancy seen in measurements of $\cl^{\kg\kcmb}$~\cite{LiuHill2015}, where the amplitude of the best-fit model compared to predictions based on \emph{Planck} CMB-derived parameters is found to be $A_{planck}=0.44\pm0.22.$\footnote{Intrinsic alignment contamination is likely to explain a significant fraction of this discrepancy, and has not been corrected for here~\cite{Chisari2015}.}  Our results using shallow galaxy samples ($i<22$ or $i<23$) are consistent with such a value, but also with $m=1$, due to the relatively large error bars. Our best-fit $m$ for the deepest sample ($i<24$) prefers a lower $m=0.6-0.7$, depending on the galaxy bias model adopted, but is also statistically consistent with a value of $m$ that would bring the CFHTLenS constraints into agreement with \emph{Planck}.  Thus, within the uncertainties of current data sets, a multiplicative shear bias remains a feasible option to reconcile the tension between the CFHTLenS and \emph{Planck} cosmological parameter constraints.  If more sensitive CMB lensing data were taken on these fields, it would be possible to improve the overall signal-to-noise such that the galaxies in the $\kgal$ reconstruction could be split into sub-samples based on different properties (e.g., color or size), perhaps allowing the cause of the multiplicative bias to be isolated.  With our current signal-to-noise, such data splits are not feasible.

As a point of comparison, we note that Ref.~\cite{Miyatake2015} compared the galaxy-galaxy lensing signal measured around SDSS luminous red galaxies using both the CFHTLenS shear catalog and the SDSS shear catalog constructed by Ref.~\cite{Mandelbaum2013}.  They found that the lensing signals agreed well, with an inverse-variance-weighted average ratio (over all radial bins) of 1.006 $\pm$ 0.046.  Since the CFHTLenS and SDSS shape measurements and photo-$z$ estimates come from completely independent pipelines, this comparison provides a constraint on any relative bias between them.  If the SDSS shear calibration were unity, then this would still leave open the possibility of a shear bias of $\approx 0.9$ for CFHTLenS (within $\approx 2\sigma$), which is consistent with the constraints on $m$ presented in this work and with the value needed to reconcile the CFHTLenS--\emph{Planck} tension.  Another possibility, albeit more unlikely, is that both catalogs have a bias in the same direction, which cancels out in the ratio of the galaxy-galaxy lensing signals measured in Ref.~\cite{Miyatake2015}.  It would be useful to perform a similar analysis to that presented in this work on the SDSS shear catalog, to independently constrain possible multiplicative biases in those data.

This study represents the first constraint on a multiplicative shear bias based on a joint cross-correlation analysis with CMB lensing data.  As our overall covariance matrix is dominated by the \emph{Planck} CMB lensing noise, galaxy lensing surveys that overlap with CMB lensing surveys with a lower noise level, e.g., the Atacama Cosmology Telescope (ACT) and the South Pole Telescope (SPT), will provide better constraints on the multiplicative bias. Furthermore, a larger sky coverage of the galaxy lensing survey will also enhance the constraint (near-future surveys are typically designed to overlap with CMB surveys). Therefore, the 5000 deg$^2$ Dark Energy Survey\footnote{\url{http://www.darkenergysurvey.org/}} (overlapping ACT and SPT), the 1500 deg$^2$ Hyper Suprime-Cam survey\footnote{\url{http://www.naoj.org/Projects/HSC/}} (entirely within ACT coverage), and the 1500 deg$^2$ Kilo-Degree Survey\footnote{\url{http://kids.strw.leidenuniv.nl/}} (overlapping ACT) will provide an excellent opportunity to study and control the multiplicative shear bias in the future.

\begin{acknowledgments}
This work would not be possible without the tremendous effort put in by the \emph{Planck} and CFHTLenS teams to make their data publicly available. We thank Ludovic van Waerbeke for providing us with binned CFHTLenS masks. We also thank Yuuki Omori, Zolt\'an Haiman, Blake Sherwin, David Spergel, and Masahiro Takada for useful discussions. We also acknowledge comments from an anonymous referee. JL is supported by National Science Foundation~(NSF) grant AST-1210877. This work was partially supported by a Junior Fellow award from the Simons Foundation to JCH. This work used the Extreme Science and Engineering Discovery Environment (XSEDE), which is supported by NSF grant ACI-1053575.

\end{acknowledgments}

\bibliographystyle{physrev}
\bibliography{main}
\end{document}